\documentclass[aps,pra,reprint,twocolumn,superscriptaddress,floatfix]{revtex4-1}
\usepackage{graphicx,amsmath,amsfonts,amssymb,amsthm,xr}
\usepackage{epsfig,amsmath,amssymb,color,dsfont,upgreek,physics}
\usepackage{mathrsfs}
\usepackage{mathtools}
\usepackage{bbold}
\usepackage{comment}
\usepackage{braket}
\usepackage{float}

\usepackage[bookmarks=true,colorlinks,linkcolor=OrangeRed,urlcolor=NavyBlue,citecolor=RoyalBlue]{hyperref}
\usepackage[dvipsnames]{xcolor}
\usepackage{orcidlink}

\definecolor{mygold}{rgb}{0.93,0.69,0.13}

\definecolor{mypurple}{rgb}{0.49,0.18,0.56}

\graphicspath{{./Figures/}}

\begin{document}
\title{Meson Mass Sets Onset Time of Anomalous Dynamical Quantum Phase Transitions}

\author{Jesse J.~Osborne${}^{\orcidlink{0000-0003-0415-0690}}$}
\email{j.osborne@student.uq.edu.au}
\affiliation{School of Mathematics and Physics, The University of Queensland, St.~Lucia, QLD 4072, Australia}

\author{Johannes Knaute${}^{\orcidlink{0000-0003-3281-1388}}$}
\affiliation{81925 Munich, Germany}

\author{Ian P.~McCulloch${}^{\orcidlink{0000-0002-8983-6327}}$}
\affiliation{Department of Physics, National Tsing Hua University, Hsinchu 30013, Taiwan}
\affiliation{Frontier Center for Theory and Computation, National Tsing Hua University, Hsinchu 30013, Taiwan}

\author{Jad C.~Halimeh${}^{\orcidlink{0000-0002-0659-7990}}$}
\email{jad.halimeh@physik.lmu.de}
\affiliation{Max Planck Institute of Quantum Optics, 85748 Garching, Germany}
\affiliation{Department of Physics and Arnold Sommerfeld Center for Theoretical Physics (ASC), Ludwig Maximilian University of Munich, 80333 Munich, Germany}
\affiliation{Munich Center for Quantum Science and Technology (MCQST), 80799 Munich, Germany}
\affiliation{Dahlem Center for Complex Quantum Systems, Free University of Berlin, 14195 Berlin, Germany}

\begin{abstract}
Dynamical quantum phase transitions (DQPTs) have been established as a rigorous framework for investigating far-from-equilibrium quantum many-body criticality. Although initially thought to be trivially connected to an order parameter flipping sign, a certain kind of \textit{anomalous} DQPTs have been discovered that exhibit no direct connection to the order parameter and have been shown to arise in the presence of confinement. Here, we show in two paradigmatic models how the onset time of anomalous DQPTs is directly connected, through a power law, to the meson mass in the confined regime of a global symmetry-broken phase. This relation becomes more prominent the closer the initial parameters are to the equilibrium quantum critical point, where a relativistic quantum field theory emerges. Our findings draw a direct connection between mesons and anomalous DQPTs, highlighting the power of the latter to classify exotic far-from-equilibrium criticality.
\end{abstract}
\date{\today} 
\maketitle

\textbf{\textit{Introduction.---}}Dynamical quantum phase transitions (DQPTs) \cite{Heyl2013,Heyl2014,Heyl2015} constitute an intuitive generalization of thermal phase transitions to far-from-equilibrium quantum many-body dynamics. In equilibrium, the partition function of a system with Hamiltonian $\hat{H}$ is $\mathcal{Z}{=}e^{{-}\beta\hat{H}}$, where $\beta$ is the inverse temperature --- we work with natural units $c{=}\hbar{=}k_\mathrm{B}{=}1$ throughout the paper. A thermal phase transition occurs at a critical inverse temperature $\beta_\mathrm{c}$, at which the free energy $\mathcal{F}{=}{-}\beta^{-1}\ln\mathcal{Z}$ exhibits a singularity \cite{Cardy_book}. In the quantum many-body context, the probability amplitude $\bra{\psi_0}e^{-i\hat{H}t}\ket{\psi_0}$, where $\ket{\psi_0}$ is an initial state \textit{quenched} by the Hamiltonian $\hat{H}$ at $t{=}0$, can be viewed as a boundary partition function where now evolution time represents a complexified inverse temperature. If one now defines the \textit{return rate} $r(t){=}{-}\lim_{L\to\infty}L^{-1}\ln\lvert\bra{\psi_0}e^{-i\hat{H}t}\ket{\psi_0}\rvert^2$, where $L$ is the system volume, this becomes equivalent to a \textit{dynamical free energy}. Singularities in $r(t)$ at critical times $t_\text{c}$ are then DQPTs, in analogy to equilibrium quantum phase transitions \cite{Heyl_review}.

Initially, DQPTs were discovered in the nearest-neighbor transverse-field Ising chain, which is integrable, and the critical times were analytically found to occur periodically \cite{Heyl2013}. Furthermore, it was shown that DQPTs only arose in this model upon quenching across the equilibrium quantum critical point. However, it was soon after shown that nonintegrable models also exhibited DQPTs and not necessarily at periodic times \cite{Karrasch2013dynamical}, and that quenching across the equilibrium quantum critical point was neither a necessary nor a sufficient condition for DQPTs to arise \cite{Vajna2014disentangling,Andraschko2014dynamical}.

DQPTs have since been studied extensively in long-range many-body models \cite{Zunkovic2016,Halimeh2017dynamical,Homrighausen2017anomalous,Zunkovic2018dynamical,Defenu2019dynamical,Halimeh2021dynamical,Corps2022dynamical,Corps2023theory,mitra2023macrostates,Corps2023mechanism,Corps2024unifying}, topological Hamiltonians \cite{Schmitt2015dynamical,Trapin2018,Sedlmayr2018bulk,Hagymasi2019dynamical,Maslowski2020quasiperiodic,Porta2020topological,Okugawa2021mirror,Sedlmayr2022dynamical,Maslowski2023dynamical}, lattice gauge theories \cite{Zache2019,Huang2019dynamical,Pedersen2021,Jensen2022,Halimeh2021achieving,VanDamme2022dynamical,Mueller2022,Pomarico2023dynamical,VanDamme2023Anatomy}, systems with broken time-translation symmetry \cite{Kosior2018dynamical,Kosior2018dynamical2}, disordered models \cite{halimeh2019dynamical,Trapin2021unconventional}, non-Hermitian systems \cite{Zhou2021,Zhou2022,Hamazaki2021Exceptional,Mondal2022Anomaly,Mondal2023Finite,mondal2024persistent}, as well as in higher spatial dimensions \cite{Bhattacharya2017emergent,Weidinger2017dynamical,Heyl2018detecting,DeNicola2019stochastic,Hashizume2022dynamical,Hashizume2020hybrid,Kosior2024vortex}, and at finite temperature \cite{Abeling2016quantum,Bhattacharya2017mixed,Heyl2017dynamical,Lang2018concurrence,Sedlmayr2018fate}. Furthermore, DQPTs have been observed experimentally in cold-atom \cite{Flaeschner2018}, trapped-ion \cite{Jurcevic2017}, and nuclear magnetic resonance \cite{Nie2020experimental} quantum simulators.

Even though it has been conjectured that DQPTs only occur in generic models when an order parameter flips sign \cite{Zunkovic2018dynamical}, this was shown not to be true \cite{Halimeh2017dynamical}. In certain models that exhibit confinement, such as long-range quantum spin chains \cite{Vanderstraeten:2018,Liu2019,Halimeh2020quasiparticle} or lattice gauge theories \cite{osborne2023probing}, so-called \textit{anomalous} DQPTs can arise that have no direct connection to the order parameter \cite{Halimeh2017dynamical}. In the quantum Ising chain with long-range interactions or a longitudinal field, domain-wall excitations can bind into mesons when the transverse-field strength is sufficiently small \cite{Kormos2017,Vanderstraeten:2018,Liu2019,Verdel2020real-time,Verdel2023dynamical,Knaute2023meson}. Similarly, in the spin-$S$ $\mathrm{U}(1)$ quantum link model (QLM) formulation \cite{Chandrasekharan1997,Wiese_review,Kasper2017} of $1+1$D lattice quantum electrodynamics (QED), a large gauge coupling \cite{Chandrasekharan1999,tang2024partialconfinementquantumlinksimulator} or a topological $\theta$-angle \cite{Surace2020,Halimeh2022tuning,Cheng2022tunable,qi2023gaugeviolationspectroscopysynthetic} can confine electron-positron pairs into mesons. In order to better understand the role of confinement in the occurrence of anomalous DQPTs, a connection between the latter and the emergent mesons is needed. 

Given that DQPTs are a promising framework for far-from-equilibrium quantum many-body universality \cite{Mori_review,Zvyagin_review}, understanding their different types and how they emerge in the presence of different dynamical phases is an essential path forward in the field. Indeed, universality classes in equilibrium phase transitions are underpinned by a set of critical exponents that are independent of the microscopic properties of the underlying models. For DQPTs to exhibit universality, it is necessary that they at least qualitatively distinguish between different dynamical phases or regimes, independently of the underlying model, with the eventual goal of extracting from them far-from-equilibrium critical exponents. Indeed, efforts in the direction of defining critical exponents due to DQPTs are already underway \cite{Heyl2014,Wu2020}.

In this Letter, we employ matrix product state (MPS) techniques \cite{Uli_review,Paeckel_review} to pin down the relation between DQPTs and mesons in the confined regimes of two paradigmatic models: the tilted Ising chain with both longitudinal and transverse fields, and the spin-$1/2$ $\mathrm{U}(1)$ QLM.

\textbf{\textit{Tilted Ising model.---}}The model we first consider here is the $1+1$D tilted Ising model \cite{McCoy1978,Fonseca2003,Coldea2010,Lake2010,Simon2011,James2019}. This model has been of particular interest in the study of confinement and false-vacuum decay \cite{Lerose2020quasilocalized,Lagnese2021,Bastianello2022,bennewitz2024simulatingmesonscatteringspin}, and its real-time confinement dynamics has recently been probed on an \texttt{IBM} superconducting processor \cite{Vovrosh2021}. It is described by the Hamiltonian
\begin{equation}\label{eq:tim}
    \hat{H}_\text{TIM} = -\sum_j\big(J \hat{\sigma}^z_j \hat{\sigma}^z_{j+1} + h \hat{\sigma}^x_j + g\hat{\sigma}^z_j\big),
\end{equation}
where \(\hat{\sigma}^x_j\) and \(\hat{\sigma}^z_j\) are the Pauli \(x\) and \(z\) matrices, respectively, acting on lattice site \(j\), \(J>0\) is the ferromagnetic coupling constant, and \(h\) and \(g\) are the transverse- and longitudinal-field strengths, respectively.
For zero longitudinal field \(g\), this model is integrable, with a phase transition at \(|h| = J\), separating a paramagnetic phase for \(|h| > J\) from a ferromagnetic phase for \(|h| < J\), which is described by the magnetization order parameter \(\hat{\mathcal{M}}^z = L^{-1} \sum_j \hat{\sigma}^z_j\), for system size \(L\) \cite{Sachdev_book}.
This ferromagnetic phase spontaneously breaks the global \(\mathbb{Z}_2\) symmetry under the transformation \(\hat{\sigma}^{z,y}_j \rightarrow -\hat{\sigma}^{z,y} _j\), \(\forall j\), and so we have a two-dimensional ground state subspace, spanned by states \(\ket{\psi^+}\) and \(\ket{\psi^-}\), with positive and negative magnetization respectively.
The lowest-energy excitations in this ferromagnetic phase are domain walls between these two ground states.

For nonzero longitudinal field \(g\), the \(\mathbb{Z}_2\) symmetry of the \(g = 0\) case is explicitly broken.
If we imagine starting in the ferromagnetic phase \(|h| < J\) with \(g = 0\) and slowly increasing \(g\), the energies of the two original ground states will start to diverge extensively, so that only one remains the true ground state, and the other becomes a sort of ``false vacuum'' \cite{Schwinger1951}.
The lowest excitations are no longer single domain walls, but a pair of domain walls confined into a \textit{meson}, being a string of false vacuum sites surrounded by the ground state.
In this sense, the longitudinal field \(g\) induces confinement in the Ising model by explicitly breaking its \(\mathbb{Z}_2\) symmetry \cite{Kormos2017}.

We consider the real-time dynamics of the tilted Ising model under a sudden quench in the longitudinal field \(g\), with a fixed value of the transverse field \(h\) just below the critical point \(h = J\): we start with the ground state with negative magnetization \(\ket{\psi^-}\) at \(g_\text{i} = 0\), and quench to \(g_\text{f} > 0\) (in this regime, this state is connected to the false vacuum, as opposed to \(\ket{\psi^+}\) which is connected to the true ground state).
We calculate this initial state and perform the time evolution using matrix product state (MPS) based numerics~\cite{Uli_review,Paeckel_review,McCulloch:2008,mptoolkit}.
We use translation-invariant MPSs, which work directly in the thermodynamic limit, find the ground state using the infinite density matrix renormalization group (iDMRG) algorithm~\cite{McCulloch:2008}, and evolve the state using the time-dependent variational principle (TDVP) for MPSs~\cite{Haegeman2016}, employing single-site updates with environment expansion~\cite{mcculloch2024comment,mcculloch2024b}.

\begin{figure}[t!]
    \includegraphics{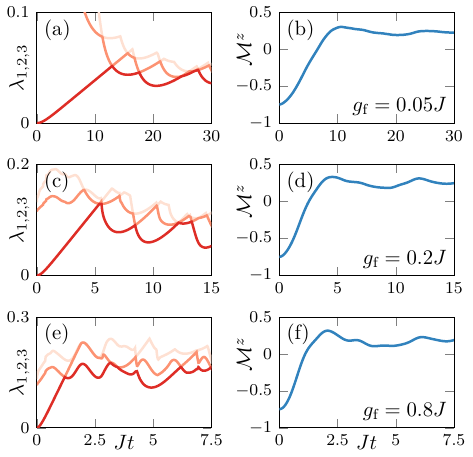}
    \caption{The evolution of the three lowest return rates \(\lambda_{1,2,3}(t)\) and magnetization \(\mathcal{M}^z(t)\) following a sudden quench in the tilted Ising model~\eqref{eq:tim} with \(h = 0.95J\), starting with the ground state with negative magnetization at \(g_\text{i} = 0\) and quenching to different longitudinal-field strengths \(g_\text{f}\).}
    \label{fig:plot-ising-h0.95}
\end{figure}

Figure~\ref{fig:plot-ising-h0.95} showcases the dynamics following this sudden quench for \(h = 0.95J\) ending at various values of \(g_\text{f}\). It displays the magnetization \(\mathcal{M}^z(t) = \ev{\hat{\mathcal{M}}^z}{\psi(t)}\), where \(\ket{\psi(t)}\) is the time-evolved state following the quench, as well as the return rates \(\lambda_n(t) = - \ln|\epsilon_n(t)|^2\), where \(\epsilon_n(t)\) is the \(n\)th eigenvalue of the MPS mixed transfer matrix \(\mathcal{T}(t)\)~\cite{Zauner2015} between the time-evolved state \(\ket{\psi(t)}\) and the initial state \(\ket{\psi^-}\), indexed by \(n\) in order of decreasing magnitude.
The overlap between these two states in the thermodynamic limit is given by the spectrum of \(\mathcal{T}(t)\) as follows,
\begin{equation}
    \lim_{L\rightarrow\infty} \braket{\psi(0)|\psi(t)} = \lim_{L\rightarrow\infty} \sum_n \epsilon_n(t)^L = \epsilon_1(t)^L,
\end{equation}
since contributions from \(n > 1\) will be exponentially suppressed: hence, we call \(\lambda_1(t)\) the return rate proper.
Nonanalyticities in \(\lambda_1(t)\), i.e., the DQPTs, will correspond to level crossings between the two highest-magnitude eigenvalues of \(\mathcal{T}(t)\).
In Fig.~\ref{fig:plot-ising-h0.95}(a,c,e), we see that, immediately following the quench, the return rate \(\lambda_1(t)\) steadily rises, until it eventually crosses over with the ``second'' return rate \(\lambda_2(t)\), signaling a DQPT, which is then followed by oscillatory behavior and more level crossings.
The magnetization, shown in Fig.~\ref{fig:plot-ising-h0.95}(b,d,f) also steadily rises from the initial value, until it changes sign and oscillates around a positive value (being the sign of the magnetization for the true ground state of the quench Hamiltonian).

\begin{figure}[t!]
    \includegraphics{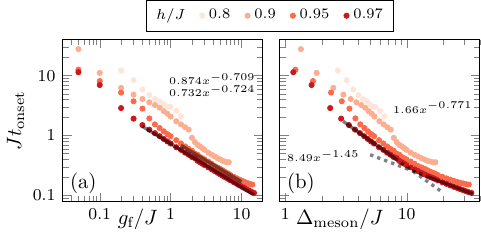}
    \caption{The onset time \(t_\text{onset}\) of the first DQPTs in the tilted Ising model~\eqref{eq:tim} following the quench protocols used in Fig.~\ref{fig:plot-ising-h0.95} for several transverse-field strengths \(h\) and a range of longitudinal-field strengths \(g_\text{f}\) following the quench.
    We show the dependence of this onset time on (a) the quench parameter \(g_\text{f}\) as well as (b) the meson gap \(\Delta_\text{meson}\) calculated at that value of \(g_\text{f}\).
    For values of \(g_\text{f}\) where the first DQPT disappears, we do not plot any data points for that \(h\).
    The superimposed lines show power-law fits, whose values are shown at the left and right of the figure (rounded to three significant figures).}
    \label{fig:onset-time-ising}
\end{figure}

As the quench value of the longitudinal field \(g_\text{f}\) is increased, the onset time \(t_\text{onset}\) of this first DQPT decreases, as shown in Fig.~\ref{fig:plot-ising-h0.95}(a,c,e).
We plot the dependence of this onset time on the quench value of the longitudinal field in Fig.~\ref{fig:onset-time-ising}(a) across a greater range of \(g_\text{f}\) for multiple values of the transverse field \(h\).
For \(h = 0.95J\) and \(0.97J\), we can clearly see a power-law relation between \(g_f\) and \(t_\text{onset}\) developing at larger \(g_\text{f}\), the lower limit of which gets smaller as we approach the critical point.
On the other hand, moving away from the critical point to smaller \(h\) (\(0.9J\) and \(0.8J\)), this linear relation starts to break down: in fact, for some ranges of the quench parameter \(g_\text{f}\), this first DQPT can disappear altogether (see the Supplemental Material \cite{SM}). It is interesting to note that in the large-entanglement regime near the critical point of this model~\eqref{eq:tim}, a relativistic quantum field theory emerges in the continuum limit \cite{Rakovszky:2016ugs,Hodsagi:2018sul}. Here, the dominant meson mass manifests as a power-law function of the longitudinal field \cite{Zamolodchikov:1989fp,Fonseca:2006au}.

Additionally, in Fig.~\ref{fig:onset-time-ising}(b), we relate the DQPT onset time \(t_\text{onset}\) with the value of the meson gap \(\Delta_{\text{meson}}\) at the value of the quench longitudinal field \(g_\text{f}\), calculated using the matrix product state excitation Ansatz~\cite{Haegeman2012}.
In this case, similarly, we can see the emergence of a power-law relation, but with a crossover in the exponent as the gap increases.
This is due to the relation between the longitudinal field and the gap, which is approximately square-root relation for small fields, but linear for larger fields \cite{SM}.
Hence, the exponent in Fig.~\ref{fig:onset-time-ising}(b) for small gaps is approximately twice the exponent of the fit in panel (a), while the exponent for larger gaps will asymptotically approach the same value as panel (a), as the offset in the aforementioned linear relation becomes negligible.

\begin{figure}[t!]
    \includegraphics{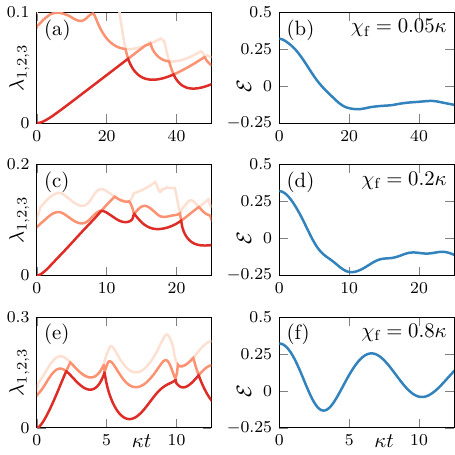}
    \caption{The evolution of the three lowest return rates \(\lambda_{1,2,3}(t)\) and electric flux \(\mathcal{E}(t)\) following a sudden quench in the quantum link model~\eqref{eq:qlm} with \(m = 0.35\kappa\), starting in the ground state with negative flux at \(\chi_\text{i} = 0\) and quenching to different confining potentials \(\chi_\text{f}\).}
    \label{fig:plot-qlm-m0.35}
\end{figure}

\begin{figure}[t!]
    \includegraphics{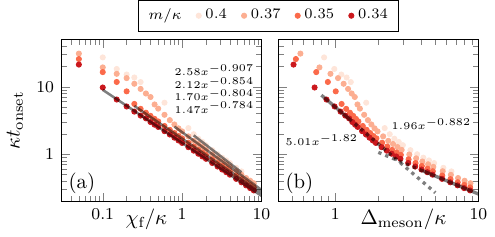}
    \caption{The same as Fig.~\ref{fig:onset-time-ising} but for the spin-$1/2$ \(\mathrm{U}(1)\) quantum link model~\eqref{eq:qlm} as the quench parameter \(\chi_\text{f}\) is tuned.}
    \label{fig:onset-time-qlm}
\end{figure}

\textbf{\textit{Quantum link model.---}}Confinement is a major phenomenon in gauge theories and a hallmark of quantum chromodynamics (QCD) \cite{Weinberg_book,Ellis_book,Peskin2016}. Recently, there has been a concerted effort to realize lattice gauge theories on quantum simulation platforms, ranging from trapped ions and cold atoms to superconducting qubits \cite{Dalmonte_review,Pasquans_review,Zohar_review,Alexeev_review,aidelsburger2021cold,Zohar_NewReview,klco2021standard,Bauer_review,funcke2023review,dimeglio2023quantum,halimeh2023coldatom}. In this vein, a quantum simulator of the spin-$1/2$ \(\mathrm{U}(1)\) quantum link model (QLM)~\cite{Chandrasekharan1997,Wiese_review} has been realized in a Bose--Hubbard quantum simulator with a tunable topological \(\theta\)-term that probed a confinement--deconfinement transition \cite{Zhang2023observation}. This motivates us to study the connection between anomalous DQPTs and the mesons in this model as well. It is described by the Hamiltonian
\begin{align}\nonumber
    \hat{H}_\text{QLM}= & -\frac{\kappa}{2} \sum_j \Big( \hat{\phi}_j^\dagger \hat{S}^+_{j,j+1} \hat{\phi}_{j+1} + \text{H.c.} \Big) \\\label{eq:qlm}
    &+ m \sum_j (-1)^j\hat{\phi}^\dagger_j\hat{\phi}_j - \chi \sum_j \hat{S}^z_{j,j+1},
\end{align}
where \(\hat{\phi}_j^\dagger\) and \(\hat{\phi}_j\) are the spinless fermion creation and annihilation operators at matter site \(j\), \(\hat{S}^\pm_{j,j+1}\) and \(\hat{S}^z_{j,j+1}\) are the spin-$1/2$ ladder and projection operators on the link between matter sites \(j\) and  \(j+1\), \(\kappa\) is the hopping strength, \(m\) is the fermionic mass, and \(\chi = g^2 (\theta-\pi) / 2\pi\) is the confining potential, with $\theta$ the topological angle.
This model has a \(\mathrm{U}(1)\) gauge symmetry generated by the local operators
\begin{equation}
    \hat{G}_j = \hat{S}^z_{j,j+1} - \hat{S}^z_{j-1,j} - \hat{\phi}^\dagger_j \hat{\phi}_j - \frac{(-1)^j-1}{2}.
\end{equation}
We work in the physical sector of states \(\ket{\psi}\) satisfying \(\hat{G}_j \ket{\psi} = 0\), \(\forall j\).

At \(\chi = 0\) (which corresponds to \(\theta = \pi\)), Coleman’s phase transition occurs at \(m_c = 0.3275\kappa\)~\cite{Coleman1976,Rico2014}, with a \(CP\)-breaking phase for \(m > m_c\).
This phase hosts a two-fold degenerate ground-state manifold spanned by ``vacuum'' states \(\ket{\psi^+}\) and \(\ket{\psi^-}\) with opposite-sign values of the electric flux \(\hat{\mathcal{E}} = L^{-1} \sum_j \hat{S}^z_{j,j+1}\).
For \(\chi \neq 0\), the particles and antiparticles feel a linear force proportional to \(\chi\) times their separation, inducing confinement: the \(CP\)-broken phase for \(m > m_c\) at \(\chi = 0\) becomes a deconfined first-order transition line between two confined phases for \(\chi > 0\) and \(\chi < 0\).

In Figure~\ref{fig:plot-qlm-m0.35}, we plot the evolution of the return rates and electric flux following a quench of the positive-flux ground state at \(m = 0.35\kappa\), \(\chi = 0\), suddenly quenching the confinement potential to a range of values \(\chi_\text{f} > 0\) (where the ground state has a negative flux).
This is similar to the quench protocol followed for the tilted Ising model in Fig.~\ref{fig:plot-ising-h0.95}, and indeed, we observe similar qualitative behavior in the return rate and order parameter, in that the onset time \(t_\text{onset}\) of the first DQPT decreases as we increase the quench parameter \(\chi_\text{f}\).
We can also see a similarity in behavior in the relation between the DQPT onset time and the quench parameter and meson gap shown in Fig.~\ref{fig:onset-time-qlm}, where again, the onset time shows an approximately power-law dependence on the quench parameter and the meson gap (although the latter again has a crossover in the exponent as we move from small to large gap, due to the relation between the gap and the confinement parameter \(\chi\) shifting from power-law to linear at larger \(\chi\) \cite{SM}).
We also can clearly see that the quality of this power-law relation gets stronger as we approach the critical point in \(m\).

Nevertheless, we can observe some differences between in the time evolution in the QLM and the tilted Ising model after this DQPT onset, such as in the strength of the oscillations of the flux in the quench to \(\chi_\text{f} = 0.8\kappa\) (Fig.~\ref{fig:plot-qlm-m0.35}(f)), and the corresponding revivals in the return rate (Fig.~\ref{fig:plot-qlm-m0.35}(e)), which we do not find in the Ising model quenches.
Whether we see strong oscillations in the order parameter or a quick equilibration to a steady value following this DQPT onset does not appear to affect the relation between the onset time and the quench parameters.

\textbf{\textit{Discussion and outlook.---}}In this Letter, we have found that the onset time of the first anomalous DQPT is determined by the confinement parameter, and also, by proxy, the meson mass, in both the tilted Ising model and the spin-$1/2$ \(\mathrm{U}(1)\) QLM with a topological \(\theta\)-term. Specifically, after instantaneous quenches of the (1+1)-dimensional tilted Ising model, the onset time decreases through a power-law as the quench value of the longitudinal field, or meson mass (gap), is increased. Analogously, we also find a power-law relation with respect to the confining potential and meson gap in the QLM. These dependencies become clearer as the equilibrium critical point is approached. In this large-entanglement regime, it is well understood that a relativistic quantum field theory emerges in the continuum limit of the Ising model (see, e.g., \cite{Rakovszky:2016ugs,Hodsagi:2018sul}), in which the dominant meson mass depends on the longitudinal field through a power law \cite{Zamolodchikov:1989fp,Fonseca:2006au}. A key finding of our studies is the observation that a similar dependence appears also for dynamical nonequilibrium quantities, independent from the underlying microscopic model. This provides a new perspective on the classification and characterization of far-from-equilibrium phenomena.

Importantly, both models considered in this work have been realized experimentally in modern quantum-simulation platforms (see, e.g., \cite{Vovrosh2021,Zhang2023observation}), several of which can also be employed to observe DQPTs. Furthermore, there have been proposals to probe meson spectra in ion-trap quantum simulators \cite{Knaue2022relativistic}, which have also been used to detect DQPTs \cite{Jurcevic2017}. In principle, one can perform meson spectroscopy by parametric modulation to search for a resonance at the meson mass. This lends hope for experimentally realizing our findings here.

An interesting future direction would be to extend our results to other more complex models, such as LGTs with non-Abelian gauge symmetries or systems in higher spatial dimensions. It would also be interesting to extract the corresponding critical exponents associated with the first anomalous DQPT through, e.g., local measures that are amenable for experimental realization \cite{Halimeh2021local}.

{\footnotesize \textbf{\textit{Acknowledgments.---}}The authors acknowledge stimulating discussions with Alvise Bastianello, Philipp Hauke, and Guo-Xian Su. J.C.H.~acknowledges support from the Max Planck Society and the Emmy Noether Programme of the German Research Foundation (DFG) under grant no.~HA 8206/1-1.
I.P.M.~acknowledges funding from the National Science and Technology Council (NSTC) Grant No.~122-2811-M-007-044.
Numerical simulations were performed on The University of Queensland's School of Mathematics and Physics Core Computing Facility \texttt{getafix}. This work is part of the Quantum Computing for High-Energy Physics (QC4HEP) working group.}

\bibliography{biblio}

\clearpage 
\pagebreak

\onecolumngrid
\begin{center}
\textbf{\large Supplemental Online Material for ``Meson Mass Sets Onset Time of Anomalous Dynamical Quantum Phase Transitions''}\\[5pt]
Jesse J.~Osborne,$^{1}$ Johannes Knaute,$^{2}$ Ian P.~McCulloch,$^{3,4}$ and Jad C.~Halimeh$^{5,6,7,8}$ \\

{\small \sl ${}^{1}$ School of Mathematics and Physics, The University of Queensland, St.~Lucia, QLD 4072, Australia}\\
{\small \sl ${}^{2}$ 81925 Munich, Germany}\\
{\small \sl ${}^{3}$ Department of Physics, National Tsing Hua University, Hsinchu 30013, Taiwan}\\
{\small \sl ${}^{4}$ Frontier Center for Theory and Computation, National Tsing Hua University, Hsinchu 30013, Taiwan}\\
{\small \sl ${}^{5}$ Max Planck Institute of Quantum Optics, 85748 Garching, Germany} \\
{\small \sl ${}^{6}$ Department of Physics and Arnold Sommerfeld Center for Theoretical Physics (ASC), Ludwig Maximilian University of Munich, 80333 Munich, Germany}\\
{\small \sl ${}^{7}$ Munich Center for Quantum Science and Technology (MCQST), 80799 Munich, Germany}\\
{\small \sl ${}^{8}$ Dahlem Center for Complex Quantum Systems, Free University Berlin, 14195 Berlin, Germany}\\
\vspace{0.1cm}
\begin{quote}
{\small In this Supplemental Material, we discuss the aberrant behavior in the DQPT onset time as we tune the confinement parameters in Figs.~\ref{fig:onset-time-ising} and \ref{fig:onset-time-qlm} in the main text, and discuss the relation between the meson gap and the confinement parameters, leading to imperfections in the power-law relations in panels (b) of the aforementioned Figures.}\\[10pt]
\end{quote}
\end{center}
\setcounter{equation}{0}
\setcounter{figure}{0}
\setcounter{table}{0}
\setcounter{page}{1}
\setcounter{section}{0}
\makeatletter
\renewcommand{\theequation}{S\arabic{equation}}
\renewcommand{\thefigure}{S\arabic{figure}}
\renewcommand{\thepage}{\arabic{page}}
\renewcommand{\thetable}{S\arabic{table}}

\vspace{0cm}

\twocolumngrid

\section*{Jumps and breaks in the DQPT onset time}
\begin{figure}[h]
    \includegraphics{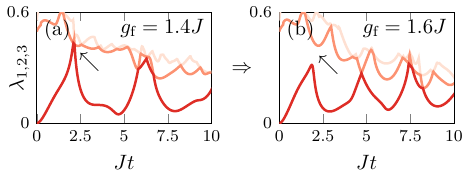}
    \caption{The evolution of the return rates \(\lambda_{1,2,3}(t)\) for tilted Ising model~\eqref{eq:tim} with \(h = 0.8J\), suddenly quenching to \(g_\text{f} = 1.4J\) and \(1.6J\), where the first DQPT disappears as \(g\) is increased.}
    \label{fig:plot-ising-break}
\end{figure}

\begin{figure}[h]
    \includegraphics{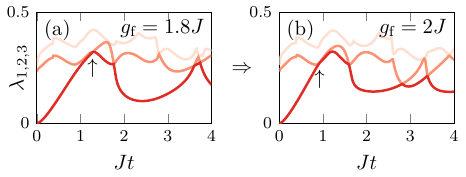}
    \caption{The evolution of the return rates \(\lambda_{1,2,3}(t)\) for tilted Ising model~\eqref{eq:tim} with \(h = 0.9J\), suddenly quenching to \(g_\text{f} = 1.8J\) and \(2J\), where the first DQPT time appears to suddenly jump in Fig.~\ref{fig:onset-time-ising}.}
    \label{fig:plot-ising-jump}
\end{figure}

When comparing the DQPT onset time as the quench parameter is tuned in Figs.~\ref{fig:onset-time-ising} and \ref{fig:onset-time-qlm}, there are some instances with sudden jumps in the onset time, or it disappearing altogether.
In Figure~\ref{fig:plot-ising-break}, we show the behavior in the return rate for \(h = 0.8J\) around the point where the onset time disappears, which is due to the first DQPT suddenly disappearing (or we could consider the DQPT at around \(Jt = 5\) at \(g_\text{f} = 1.6J\) to be the ``new'' initial DQPT, in which case this break in the onset time would become a discontinuous jump to a higher value).
In Figure~\ref{fig:plot-ising-jump}, we look at \(h = 0.9J\) around the point where there appears to be sudden jump in the onset time in Fig.~\ref{fig:onset-time-ising}, which is due to the intersection of the two lowest return rates suddenly moving from the peak to the trough of the second return rate.
As we approach the critical point, these jumps and breaks become less prevalent, and the relation between the onset time and quench parameter more closely follows a power law.

\section*{The relation between the meson gap and the confinement parameter}
\begin{figure}[h]
    \includegraphics{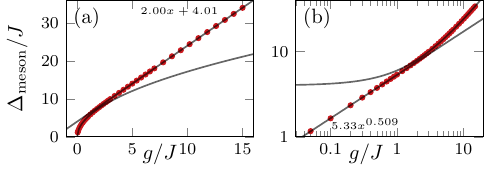}
    \caption{The relation between the meson gap and the longitudinal field \(g\) in the tilted Ising model~\eqref{eq:tim} with \(h = 0.97J\) on (a) a linear and (b) a log–log scale.
    The superimposed lines show a power-law and linear fit for small and large \(g\), respectively, with the values given next to the relevant fit (rounded to three significant figures).}
    \label{fig:ising-gap}
\end{figure}

\begin{figure}[h]
    \includegraphics{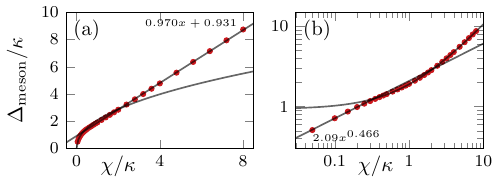}
    \caption{The same as Fig.~\ref{fig:ising-gap}, but for the \(\mathrm{U}(1)\) quantum link model \eqref{eq:qlm} at \(m = 0.34\kappa\), as the confinement parameter potential \(\chi\) is tuned.}
    \label{fig:qlm-gap}
\end{figure}

In Figure~\ref{fig:ising-gap}, we show the relation between the meson gap and the longitudinal-field strength \(g\) in the tilted Ising model~\eqref{eq:tim}.
We can see a clear crossover in behavior here: For small \(g\), the meson gap follows a power-law relation \(Ag^\alpha\), while for larger \(g\), the meson gap follows a linear relation \(mg + c\).
This crossover also appears in the QLM as the confinement potential \(\chi\) is tuned, as shown in Fig.~\ref{fig:qlm-gap}.
Because of this, the plots of the meson gap versus the confinement parameters in Figs.~\ref{fig:onset-time-ising}(b) and \ref{fig:onset-time-qlm}(b) appear slightly bent, and do not exactly follow a power-law relation.
\end{document}